\documentclass[epsf]{emulateapj}
\shorttitle{Globular Cluster Metallicity Bimodality}
\shortauthors{Brodie \etal~}
\def\etal{{\it et al.}}

\usepackage{subfigure}
\usepackage{graphics}
\begin{document}

\title{The SLUGGS Survey: NGC 3115, A Critical Test Case for Metallicity Bimodality in Globular Cluster Systems}

\author{Jean P.~Brodie\altaffilmark{1}, Christopher Usher\altaffilmark{2}, Charlie Conroy\altaffilmark{1}, Jay Strader\altaffilmark{3}, Jacob A. Arnold\altaffilmark{1}, Duncan A.~Forbes\altaffilmark{2}, Aaron J.~Romanowsky\altaffilmark{1,4}}
\email{brodie@ucolick.org}

\affil{
\altaffilmark{1}University of California Observatories, 1156 High Street, Santa Cruz, CA 95064, USA\\
\altaffilmark{2}Centre for Astrophysics \& Supercomputing, Swinburne University, Hawthorn, VIC 3122, Australia\\
\altaffilmark{3}Department of Physics and Astronomy, Michigan State University, East Lansing, Michigan 48824, USA\\
\altaffilmark{4}Department of Physics and Astronomy, San Jos\'e State University, One Washington Square, San Jose, CA 95192, USA\\}
\begin{abstract}

Due to its proximity (9 Mpc) and the strongly bimodal color distribution of its spectroscopically well-sampled globular cluster (GC) system, the early-type galaxy NGC 3115 provides one of the best available tests of whether the color bimodality widely observed in GC systems generally reflects a true metallicity bimodality. Color bimodality has alternatively been attributed to a strongly nonlinear color--metallicity relation reflecting the influence of hot horizontal branch stars. Here we couple Subaru Suprime-Cam $gi$ photometry with Keck/\textsc{deimos} spectroscopy to accurately measure GC colors and a CaT index that measures the  \ion{Ca}{2} triplet. We find the NGC 3115 GC system to be unambiguously bimodal in both color and the CaT index. Using simple stellar population models, we show that the CaT index is essentially unaffected by variations in horizontal branch morphology over the range of metallicities relevant to GC systems (and is thus a robust indicator of metallicity) and confirm bimodality in the metallicity distribution. We assess the existing evidence for and against multiple $\it{metallicity}$ subpopulations in early and late-type galaxies and conclude that metallicity bi/multimodality is common. We briefly discuss how this fundamental characteristic links directly to the star formation and assembly histories of galaxies.

\end{abstract}


\section{Introduction}

The fact that most massive galaxies possess globular cluster (GC) systems that are bimodal in color has been recognized for over a decade (e.g., \citealt{1993MNRAS.264..611Z, 1993AJ....105.1762O}). The general presence in galaxy GC systems of two principal subpopulations is implicit in this finding.
It is now well-established that the vast majority of GCs are old ($\ga 10$ Gyr; e.g., \citealt{2005AJ....130.1315S, 2005A&A...439..997P}),
which suggests that the subpopulations are separated predominantly by metallicity rather than age.

Considerable effort has since been devoted to understanding the origin of metallicity bimodality as it has significant implications for understanding the star formation and assembly histories of galaxies (see review by \citealt{2006ARA&A..44..193B}). This follows from the notion that GC formation accompanies all major star formation events in a galaxy's history.  Once formed, GCs remain as unchanging (compared to galaxy starlight) bright beacons, and are the accompanying witnesses to the development of cosmic structure.  

This general view has been questioned, however. Richtler (2006), showed that a hypothetical unimodal metallicity distribution function (MDF) can appear bimodal in some optical colors when scatter 
is included in the color-metallicity relation. \citet{2006Sci...311.1129Y} suggested, based on their single stellar population (SSP) models, that a strong inflection appears in the central part of the optical color--metallicity relations (between [$Z$/H]= $-1.5$ and $-0.5$), driven by contributions from horizontal branch (HB) stars. This inflection can transform a unimodal MDF into two color peaks in their models. Recently, \citep{2011ApJ...743..149Y, 2011ApJ...743..150Y} explored their expectations for ultraviolet/optical color combinations and concluded, again, that the underlying GC MDFs are generally unimodal. 
Recent investigations have found evidence for and against metallicity bimodality (\citealt{2011MNRAS.417.1823A, 2010AJ....139.1566F, 2012A&A...539A..54C, 2012ApJ...746...88B}) without a clear resolution of the issue.

The most obvious objection to the idea of a unimodal MDF for GCs is that it conflicts with our knowledge of the best-studied GC system of all, that of the Milky Way. Here metallicities are tied to accurate abundance measurements for individual stars and the unambiguously bimodal metallicity distribution links directly to distinct spatial and kinematical GC subsystems \citep{SZ78}. 
Nonetheless, a generalization to other galaxies must rely on more limited information, such as metallicities derived from integrated spectroscopy, as we cannot obtain full chemical and phase space data for systems beyond the Local Group (although see \citealt{Pota12} for the results of a color--kinematical survey of nearby GC systems).

In principle, to resolve the bimodality question in a convincing way we would need to identify galaxies having the most clearly bimodal color distributions and derive metallicities from spectra for a significant fraction of their GC subpopulations. 
Galaxies with unclear color bimodality tell us little about this particular issue, although they may be revealing about the complexities of galaxy assembly, as discussed below. If, in even one unequivocal case, a bimodal fit to the MDF were to be strongly preferred over a unimodal one, and the proportion of metal-rich to metal-poor GCs were to be similar to the red/blue number ratio, this would provide convincing support for the principle of distinct GC metallicity subpopulations in galaxies. Similarly, if even one case with strong color bimodality turned out to be
unambiguously unimodal in metallicity, this would provide a huge boost to the unimodal metallicity hypothesis.

Here we present new spectroscopically-based measures of GC metallicity in a nearby galaxy, NGC 3115, that offers the best available test in an early-type galaxy of these distinct interpretations of color bimodality.  NGC 3115, one of 11 nearby galaxies in the spectroscopic study of \citet{Usher12}, is an S0 galaxy hosting a populous GC system that is strongly bimodal in optical colors. 

The great advantage of NGC 3115 for exploring the question of metallicity bimodality is 
its proximity ($\sim 9.4$ Mpc; \citealt{2001ApJ...546..681T}). This is important because spectroscopic studies that generate reliable metallicity estimates for GCs are very challenging, even for the largest telescopes equipped with efficient spectrographs. This has led to the tendency for spectroscopically-derived metallicity estimates to be confined to the brighter subset of GCs in all but the closest galaxies. The brightest GCs are liable to display unimodal color and metallicity distributions because of the ``blue tilt" phenomenon, or mass-metallicity relation, that reflects self-enrichment for massive metal-poor GCs.\citep{2006AJ....132.2333S, 2006ApJ...636...90H, 2006ApJ...653..193M}. The blue tilt tends to reduce the separation of the red and blue distributions with increasing brightness in the color--magnitude plane.

For NGC 3115, which has a populous GC system consisting of $546\pm80$ GCs within 50 kpc \citep{2011MNRAS.416..155F}, we spectroscopically sample a significant fraction ($\sim$20\%) of the ordinary GC population. Indeed, we obtain a reliably measured metallicity index, independent of both color and SSP modeling considerations, down to the turnover of the GC luminosity function.

Section 2 describes the new observations that allow us to characterize the metallicity distribution of NGC 3115 GCs. Section 3 presents colors and measurements of the metallicity-sensitive \ion{Ca}{2} triplet index (CaT) for these GCs, and our investigation of the dependence of the CaT on HB morphology. Section 4 is a discussion of these findings in the context of arguments in the literature for and against metallicity bimodality. Section 5 contains the summary and conclusions. 

\section{Observations}\label{sec:obs}

To study the metallicity distribution of the NGC 3115 GC system we use the Subaru Suprime-Cam $gi$ photometry and the Keck/\textsc{deimos} multiobject spectroscopy given by \citet{2011ApJ...736L..26A} but with the color corrections and revised zero points presented in \citet{Usher12}. These data were taken as part of the SLUGGS survey\footnote{http://sluggs.ucolick.org}. The spectroscopic program was designed to target the region around the Ca II triplet at 8498, 8542 and 8662 \AA\ (CaT). To negate the effect of the strong sky lines in this spectral region we employ a technique described in detail in \citet{2010AJ....139.1566F, 2011MNRAS.415.3393F} and fit a linear combination of stellar template spectra to the observed spectra while masking out those spectral regions containing significant sky lines. The fitted spectra are then continuum-normalized and the strength of the CaT index is measured from the normalized spectra. A Monte Carlo resampling technique is used to derive asymmetric confidence intervals for each measurement. Further details of the CaT observations and measurement techniques are given in \citet{Usher12}.  

\section{GC Metallicities}\label{sec:mets}

CaT has long been known to correlate with metallicity for old stellar populations (e.g., \citealt{1988AJ.....96...92A, 2002MNRAS.329..863C}) but issues have been raised about its usefulness as a one-to-one predictor of metallicity because it may become less sensitive (``saturate") at higher metallicities ([$Z$/H]$> -$0.4) and because it may respond to changes in HB morphology due to the presence of Paschen lines in the line index bandpasses \citep{2010AJ....139.1566F}.  SSP modeling now shows that the flattening at high metallicities is mild, and that the effect of the HB is minimal (see below). 

Figure 1 is a plot of  the highest S/N ($>$12 \AA\ $^{-1}$) CaT measurements  for 71 GCs in NGC 3115 as a function of $(g-i)_0$ color, along with the corresponding histograms. A cut of  ($> $8 \AA\ $^{-1}$) increases the sample to 122 but does not significantly change the robustness of the Gaussian Mixture Modeling (GMM) statistics \citep{2010ApJ...718.1266M}. This galaxy exhibits the clearest color bimodality of any known GC system ($>99.9$\% confidence). Its DD value (a GMM measure of peak separation divided by peak widths) is 3.9. By comparison,  bimodality ``poster children" NGC 4594 and NGC 4472 have DD values of 3.2 ($B-R$) and 3.0 ($g-z$), respectively. Over-plotted in the upper panel of Figure 1 is the color histogram for the entire NGC 3115 photometric sample, showing that the spectroscopically observed subset faithfully represents the overall GC distribution. The bimodal nature of both the color and the CaT index distributions is unambiguous.

\begin{figure}
\includegraphics[width=\columnwidth]{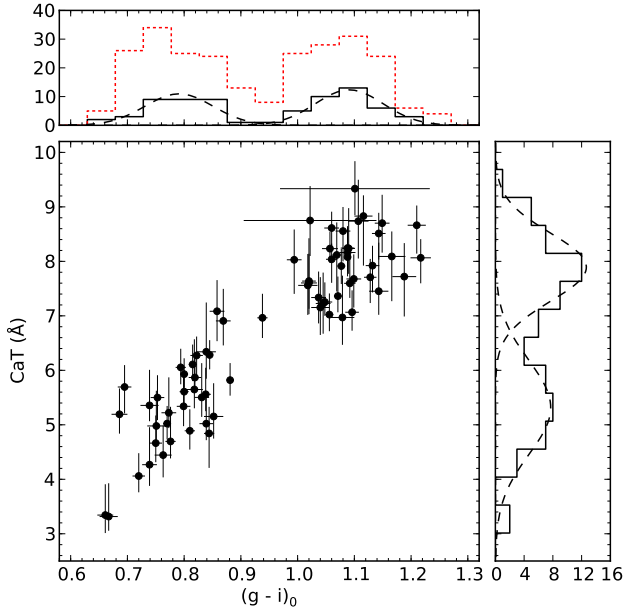}
\caption{CaT index measurements versus $(g-i)_0$.
The solid and dotted histograms are the spectroscopic (71 GCs) and the entire, background decontaminated, photometric (253 candidates) samples respectively. The overall color distribution is well-represented by the spectroscopic sample. Overplotted on the histograms are the best-fit Gaussians from a standard mixture-modeling analysis. Bimodality is preferred over unimodality at the $>99.9$\% and 99.8\% levels for the color and CaT index distributions respectively.}
\end{figure}

Possible index sensitivity to the distribution of stars on the HB is investigated by measuring CaT in model spectra, using the same methods used for GC measurements.
The models are based on those of Conroy and collaborators \citep{2009ApJ...699..486C, 2010ApJ...712..833C}, but have new synthetic spectral libraries to allow for $\alpha$-enhancement and explicit modeling of the CaT index. 
Two extreme cases are investigated in a 13 Gyr single stellar population, one with an extreme blue HB at all metallicities and the other with an all-red HB. Figure 2 demonstrates conclusively that the CaT index is effectively insensitive to HB morphology; while the spectra themselves change with the addition of hot stars, the effects are present in both the index bandpass and pseudocontinua and largely cancel out. Figure 2 also shows that $\alpha$-enhancement does not change the shape of the relation between metallicity and the CaT index, as the enhanced models simply move along the relation defined by the solar-scaled models. CaT  is thus an accurate measure of total metallicity, [$Z$/H], in GCs (where [$Z$/H]=[Fe/H] + 0.934 $\times$ [$\alpha$/Fe]). The relation between CaT and [$Z$/H] is approximately linear and permits excellent metallicity discrimination over essentially the full range of observed GC metallicities (the models span $-2.0<$[$Z$/H]$<0.0$).

\begin{figure}
\includegraphics[width=\columnwidth]{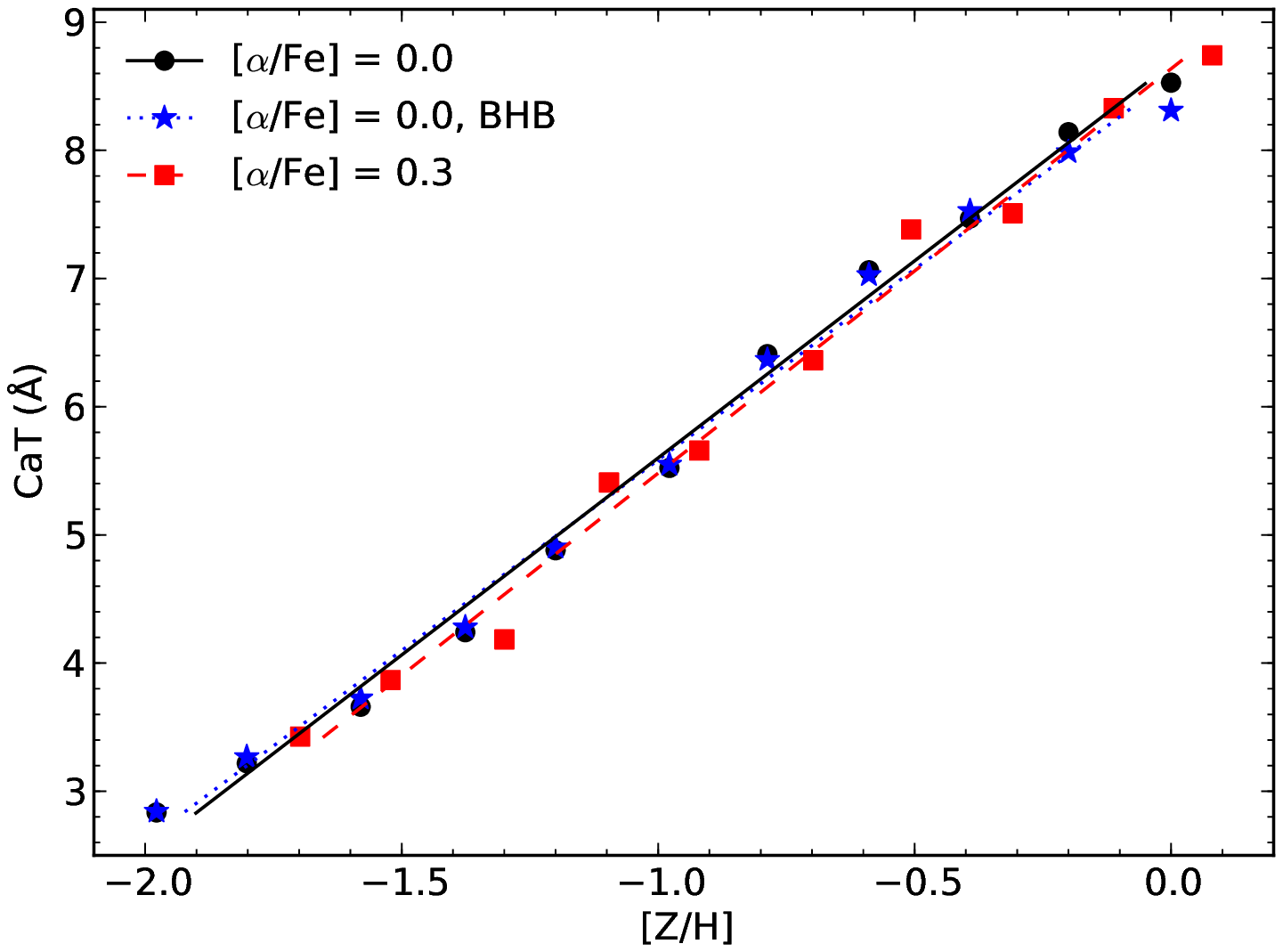}
\caption{CaT index versus metallicity based on the 13 Gyr SSP models of \citet{2009ApJ...699..486C, 2010ApJ...712..833C}. Dotted line/star symbols are a model with an extreme blue HB at all metallicities and no $\alpha$-enhancement. Solid line/filled circles represents a purely red HB population, again with solar [$\alpha$/Fe]. The dashed line/square symbols have [$\alpha$/Fe]=0.3. Although blue HB stars do influence the Paschen lines, their effect on the CaT index is negligible. Our CaT measurements are insensitive to [$\alpha$/Fe].}
\end{figure}

Figure 3 is a plot of [$Z$/H], derived from CaT measurements using the CaT--[$Z$/H] relation of \citet{Usher12}, against  $(g-i)_0$ color. 
Usher \etal~compared data and SSP models for a large number of galaxies' GC systems and defined a robust linear relation between CaT and [$Z$/H]. A GMM
analysis finds the [$Z$/H] distribution of the NGC 3115 GCs to be bimodal at the 99.8\% confidence level. 
Overplotted are the \citet{2011ApJ...743..150Y} color--metallicity relation and a two--component linear fit to the NGC 3115 data.
The broken linear fit provides a better representation of the data. A similar result was found for M31 GCs, where  \citet{2011ApJ...737....5P} used $ugriz$ optical photometry and spectroscopic metallicities and found color--metallicity relations that are all approximately linear, without the sharp inflections predicted by the \citet{2011ApJ...743..150Y} models.

\begin{figure}
\includegraphics[width=\columnwidth]{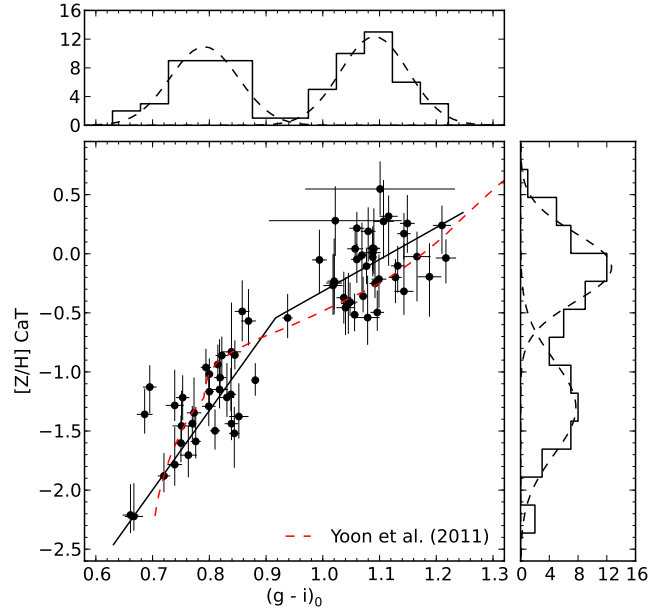}
\caption{GC color--metallicity relation. Metallicities are derived from our CaT measurement for 71 GCs, using the CaT--[$Z$/H] relation of \citet{Usher12}.  The dashed line is the color--metallicity model relation of \citet{2011ApJ...743..150Y} and the solid line is a two--component, best-fit linear color--metallicity relation that provides a better representation of the data.}
\end{figure}

Figure 4 presents a comparison of the MDF derived from the CaT index with MDFs obtained by applying the \citet{2011ApJ...743..150Y} model, and the \citet{Usher12} empirical color--[$Z$/H] relation (their equation 10), to the observed $(g-i)_0$ colors. While the Yoon \etal~color--[$Z$/H] relation appears to provide an acceptable (although not optimal) fit to the [Z/H] data in Figure 3, it produces an MDF that is single-peaked with an extended metal-poor tail. This differs dramatically from the bimodal MDFs derived spectroscopically and from the \citet{Usher12} relation.

\begin{figure}
\includegraphics[width=\columnwidth]{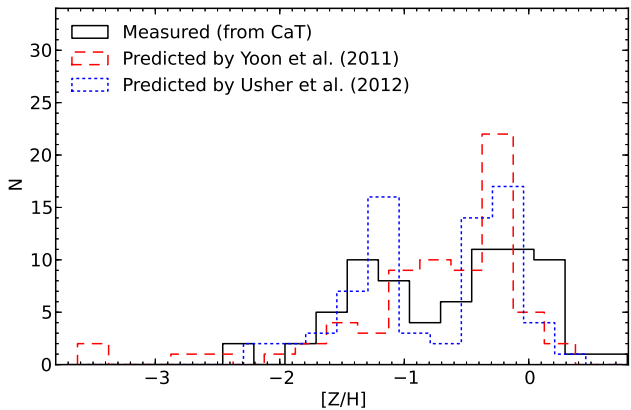}
\caption{GC metallicity distribution functions. The solid line is the MDF derived from CaT measurements. It is bimodal at the 99.8\% confidence level.
The dashed and dotted lines are obtained from $(g-i)_0$ colors using the \citet{2011ApJ...743..150Y} model color--[$Z$/H] relation, and the \citet{Usher12} empirical color--[$Z$/H] relation respectively.} 
\end{figure}

\section{Discussion}\label{sec:disc}

Here we assess the cumulative evidence for and against GC metallicity bimodality and briefly discuss the relevance of GC metallicity subpopulations in unraveling the assembly histories of galaxies.  

\subsection{Optical Spectroscopy}\label{sec:spectra}

Apart from the obvious example of our own Milky Way, where metallicity bimodality is irrefutable,  there are several galaxies for which a reasonably large sample of GC metallicities have been measured directly from their spectra. M49 (NGC 4472), the brightest galaxy in the Virgo cluster, was studied by \citet{2007AJ....133.2015S}  who used the Keck/LRIS spectroscopy of \citet{2003ApJ...592..866C} to estimate metallicities from Lick indices for 47 of its GCs. The resulting MDF was shown to be bimodal, and the locations of the red and blue peaks were shown to be consistent with expectations based on the GC metallicity--host galaxy mass relation
\citep{2006AJ....132.2333S, 2006ApJ...639..838P}.   

More recently, \citet{2011MNRAS.417.1823A} used 112 high quality Keck/DEIMOS spectra of confirmed GCs in the Sombrero galaxy (NGC 4594) to measure metallicities via the \citet{1990ApJ...362..503B} method. Again, the MDF is clearly bimodal (KMM probability $>$90\%). Moreover, they find evidence for a linear transformation between metallicity and $(B-R)$ color.

The main weakness of the NGC 4472 work is the small sample size, especially as a fraction of the total GC system, and that only the bright end of the luminosity function is sampled. While the sample in NGC 4594 is larger, the spectra are generally of lesser signal-to-noise than those of the 71 selected objects in NGC 3115.  More critically, the NGC 3115 system is the definitive litmus test of metallicity bimodality because its color bimodality is much clearer than in either NGC 4472 or NGC 4594.

NGC 5128 (Cen A) is another galaxy with a well-studied GC system.  \citet{2010ApJ...708.1335W} found bi/multimodality in the spectroscopically derived metallicities and colors of the old GCs in their sample. However, NGC 5128 is known to be the remnant of a recent merger and has a population of relatively young GCs, presumed to have formed in that merger \citep{2008MNRAS.386.1443B}. Although this galaxy is GC-rich and nearby, its complexity and special, post-merger status make it far from ideal as a bimodality benchmark. 

Two well-studied GC-populous galaxies, M31 \citep{2011AJ....141...61C} and M87 \citep{1998ApJ...496..808C}, have spectroscopic metallicity distributions that are not clearly bimodal, and for these the issue is not with sample size or signal-to-noise. Neither of these galaxies would fit our criterion for clear color bimodality. The peaks of the color distributions are poorly separated and the kurtoses are positive (M31: $DD = 1.47$, $k = 0.23$; M87: $DD = 1.30$, $k = 0.19$). Therefore, the lack of metallicity bimodality is to be expected. Moreover, only GCs significantly brighter than the mean were included in the Cohen \etal~study.  Interestingly though, in M87, \citet{2011ApJS..197...33S} found evidence from subsample kinematics and colors that multiple, old populations may be present, perhaps as a result of ongoing accretion of infalling satellite galaxies hosting GC systems with a range of masses and hence colors and metallicities. In general, the extended formation histories of central cluster galaxies like M87 may be expected to produce more complex GC systems, especially when studied in detail. M31 is also actively accreting satellite galaxies and their associated GCs \citep{2010ApJ...717L..11M}, although it is not clear whether this recent accretion history is typical of or unusual for massive disk galaxies.

An important point in the context of this work is that spectroscopically inferred MDFs, whether unimodal or multimodal, do not appear to be well-reproduced by application of the Yoon \etal~models. 
For example, in M31, using the photometry of \citet{2010MNRAS.402..803P} and the spectroscopic metallicities of \citet{2011AJ....141...61C}, we find that the KS test gives a probability of 4 $\times 10^{-8}$ that the metallicities predicted by the Yoon \etal~ models are drawn from the same distribution as as the spectroscopic metallicities. A generic result from the Yoon \etal~models is that they produce a single skewed metallicity distribution with a broad peak, steeply declining at the metal-rich end and with a long tail to metal-poor clusters.  \citet{Usher12} show that the observed [$Z$/H] GC distributions  for 11 early-type galaxies are not as skewed or peaky as the Yoon \etal~models would predict.

\subsection{Near Infrared Colors}\label{sec:NIR}

Near-infrared colors, primarily reflecting the temperatures of stars on the red giant branch, should be less sensitive to the distribution of stars on the HB. Thus a unimodal metallicity distribution should map into a unimodal near-infrared color distribution, and the detection of bimodality in a near-infrared color distribution would strongly imply an underlying bimodal metallicity distribution. \citet{2012A&A...539A..54C} argued that $K$-band photometry is key to a clean discrimination of metallicity, but their detailed modeling of photometric scatter showed how difficult it is to obtain the sample size and data quality needed to reveal the effect. The difficulty is underscored in M87, where \citet{2007ApJ...660L.109K} found clear GC color bimodality in $I-H$, based on NICMOS observations, while \citet{2012A&A...539A..54C} failed to find the effect in $g-K$ with a larger sample. 

Although NGC 1399 shows significant GC bimodality in optical colors, \citet{2012ApJ...746...88B} found it to be unimodal in $I-H$. 
A worry is that their inferred MDF does not seem to be consistent with results from the actual spectroscopic MDFs of other galaxies, as these consistently have much more significant metal-poor subpopulations than the MDFs produced from photometry using nonlinear color--metallicity relations.  Spectroscopically-derived metallicities are urgently needed for NGC 1399 GC system.
Chies-Santos \etal~found that the $g-K$ color distribution of NGC 4649 (M60) is best described by a bimodal distribution (KMM probability $>$95\%), consistent with metallicity bimodality in this galaxy's GC system. Using Spitzer observations, \citet{2008MNRAS.389.1150S} also found bimodal optical/mid-IR (3.6 microns) color distributions in the GC systems of NGC 5128 and NGC 4594.

\subsection{Multiple GC Subpopulations and Galaxy Formation}\label{sec:galform} 

The well-established correlations between GC color and host galaxy properties \citep{1997AJ....113.1652F, 2001AJ....121.2974L, 2006AJ....132.2333S, 2006ApJ...639..838P} must be linked directly to metallicity effects to establish GCs as chemodynamical tracers of the star formation and assembly processes that shaped galaxies over cosmic time.

Several lines of evidence, including spatial and metallicity distributions and specific frequencies, have led to suggestions that metal-poor blue GCs are formed in dark matter halos at the earliest times and are associated with metal-poor galaxy halos \citep{2005AJ....130.1315S, 2012arXiv1205.5315F}, while the metal-rich red GCs closely trace the build up of galaxy bulges \citep{2001ApJ...556L..83F, 2005AJ....130.1315S, Pota12}.  

The observed bimodality might arise naturally in the hierarchical merging paradigm as a consequence of the fundamental relation between GC metallicity and host galaxy mass \citep{2010ApJ...718.1266M}, or it may be the result of a temporary truncation of GC formation at high redshift \citep{2006ARA&A..44..193B}.

Regardless of the eventual viability of these scenarios, the linkage of GC and galaxy properties can be expected to provide fundamental, unique
constraints on galaxy formation in a hierarchical merging framework.

\section{Summary and Conclusions}\label{sec:sumconc}

CaT indices and $gi$ colors have been measured for 71 GCs with high signal-to-noise spectra in NGC 3115, a nearby S0 galaxy with a populous GC system that exhibits the most strongly bimodal color distribution of any known GC system. The CaT index has been shown to be effectively insensitive to HB morphology and $\alpha$-enhancement 
over a range of metallicities from [$Z$/H]=$-2.0$ to [$Z$/H]=0.0 using new stellar population models. CaT correlates closely with [$Z$/H] over the range of interest for GCs; it is an excellent proxy for metallicity in such systems. Color and CaT (and hence metallicity) are shown to be bimodal for NGC 3115 GCS with high statistical significance ($>99.9$\% and 99.8\%, respectively).

The suggestion of \citet{2006Sci...311.1129Y}  that the underlying metallicity distributions of GC systems are unimodal, and that the observed color bimodality is an artifact of strongly nonlinear color--metallicity transformations, is not confirmed in this galaxy. While this result does not {\it necessarily} apply to all early-type galaxies, it does lend considerable weight to the mounting evidence from direct MDFs that multiple GC metallicity subpopulations are common. The fact that a variety of optical colors and the easily-measured CaT index are excellent indicators of metallicity provides us with a key diagnostic for unravelling the star formation and assembly histories of galaxies.

\acknowledgments

Data presented herein were obtained at the W.~M.~Keck Observatory, which is operated as a scientific partnership among Caltech, UC, and NASA. Based on data collected at Subaru (operated by NAOJ) via Gemini time exchange (GN-2008A-C-12). This work was supported by the NSF through grants AST-0808099, AST-0909237, and AST-1109878. \\

\end{document}